\documentclass[12pt]{article}
\pdfoutput=1
\usepackage{graphicx,psfrag,epsf,color}
\usepackage{xcolor}
\usepackage{amsmath,amssymb,amsfonts}
\usepackage{url}
\usepackage{multirow}
\usepackage{geometry}
\usepackage{float}
\usepackage{mathrsfs}  
\usepackage{authblk}
\usepackage{array}
\usepackage{cite}
\usepackage{slashed}
\usepackage{graphicx}
\usepackage{subcaption}
\usepackage{rotating}
\usepackage{slashed, cancel}
\usepackage{placeins}
\usepackage{rotating}
\usepackage{soul}
\usepackage{booktabs}
\usepackage{hyperref}

\newcounter{MBQ}

\newcounter{MBQQ}

\usepackage{tabularx}
\newcolumntype{C}{>{\centering\arraybackslash}X}

\definecolor{ForestGreen}{RGB}{34,139,34}

\usepackage{bigints}

\newcommand{\DC}{\Delta C}

\def\be{\begin{equation}}
\def\ee{\end{equation}}
\def\beq{\begin{eqnarray}}
\def\eeq{\end{eqnarray}}
\usepackage{hyperref}

\def\app#1{{App.~\ref{#1}}}

\newcommand{\sLb}{s_{\Lambda_b}}
\newcommand{\sL}{s_\Lambda}
\numberwithin{equation}{section}

\begin{document}
\allowdisplaybreaks

\begin{titlepage}

\begin{flushright}
{\small
ZU-TH-76/25\\
\today \\
}
\end{flushright}

\vskip0.8cm
\begin{center}
{\Large \bf\boldmath 
Probing  Lepton Flavour Universality\\[3pt]
with $\Lambda_b$ decays to $\tau^+\tau^-$ final states
}
\end{center}

\vspace{0.5cm}
\begin{center}
{\sc Marzia~Bordone,$^{a}$ Gino~Isidori,$^a$ \\
Christiane~Mayer,$^{a}$ and Jan-Niklas Toelstede$^{a,b}$} \\[6mm]
{{\it $^a$Physik-Institut, Universit\"at Z\"urich, CH-8057 Z\"urich, Switzerland}\\
{\it $^b$Paul Scherrer Institut, Forschungsstrasse 111, CH-5232 Villigen, Switzerland}}
\end{center}

\vspace{0.6cm}
\begin{abstract}
We present a study of the rare baryonic decay $\Lambda_b \to \Lambda \tau^+ \tau^-$ as a probe of new physics (NP) coupled preferentially to third-generation fermions. Within the Standard Model, we evaluate the branching ratio and the lepton-flavour-universality (LFU) ratio $R_{\Lambda}^{\tau/\mu}$, including both perturbative and long-distance charm contributions. We show that the LFU ratio can be predicted with an uncertainty below 10\%. Possible NP effects arising from lepton non-universal dynamics are analysed within an effective field theory framework motivated by the current anomalies in $b \to c\tau\nu$ and $b \to s\mu^+\mu^-$ transitions. In this context, $R_{\Lambda}^{\tau/\mu}$ can be enhanced by several orders of magnitude, offering a clear target for upcoming searches. The implications for the related mode $\Lambda_b \to pK\tau^+\tau^-$ are also briefly discussed.
\end{abstract}
\end{titlepage}


\section{Introduction}
\label{sec:intro}

One of the most interesting and well-motivated hypothesis about the ultraviolet completion of the  Standard Model (SM) is that it consists of new degrees of freedom coupled predominantly to third-generation fermions, at least at nearby energy scales.
This first layer of New Physics (NP) should also be coupled weakly and in an almost flavour-universal manner to the lighter families. 
This general setup is supported by the observed fermion mass hierarchies and  
the electroweak hierarchy problem, while being consistent with the absence of significant deviations in flavour-changing processes and electroweak precision tests~\cite{Davighi:2023iks}.
Direct searches at the LHC are also considerably weaker for new states 
coupled mainly to third-generation fermions, which can still be as light 
as 1--2~TeV~\cite{Allwicher:2023shc}.

In order to test this hypothesis via flavour observables, a particularly interesting role is played by the flavour-changing neutral current (FCNC) transitions of the type $b \to s \tau^+ \tau^-$.
Experimentally, these modes are extremely challenging to access, and 
current bounds are still several orders of magnitude above the 
SM expectations.  Nevertheless, the potential  NP 
effects in these channels can be very large:
in well-motivated scenarios, the enhancement over SM rates may reach up to three orders of magnitude (see e.g.~Ref.~\cite{Alonso:2015sja,Capdevila:2017iqn,Kumar:2018kmr,Cornella:2019hct,Cornella:2021sby}). 
Recent improvements in sensitivity from the LHCb~\cite{LHCb:2025lcw} and 
Belle~II~\cite{Belle-II:2025lwo} collaborations have begun to open this window, with future prospects appearing especially promising. 
Even if the SM regime remains out of reach, the expected progress could soon provide meaningful constraints on NP models that predict large effects in these modes.

From a phenomenological point of view, the study of 
$b \to s \tau^+ \tau^-$ transitions is further motivated by the 
persistent discrepancies between data and SM predictions 
observed in both $b \to s \mu^+ \mu^-$ and $b \to c \tau \nu$ transitions. 
In the latter case,  a clear correlation between charged- and neutral-current processes is dictated by the electroweak symmetry. As a result of this correlation,  the Lepton Flavour Universality (LFU) ratios $R_D$ and $R_{D^*}$  imply  large enhancements in the $b \to s \tau^+ \tau^-$ rates,
if interpreted as a NP signals~\cite{Alonso:2015sja,Capdevila:2017iqn,Kumar:2018kmr,Cornella:2019hct,Cornella:2021sby}. Moreover, the $b \to s \tau^+ \tau^-$ amplitude thus predicted 
radiatively induce a smaller deviation in $b \to s \ell^+ \ell^-$ 
($\ell=e,\mu$)~\cite{Bobeth:2011st,Crivellin:2018yvo,Alguero:2022wkd}
whose magnitude and size is in the right ballpark to explain current tensions in various $b \to s \mu^+ \mu^-$ observables.

In this work, we focus on a specific exclusive $b \to s \tau^+ \tau^-$ process 
that appears particularly promising for experimental investigation at hadron colliders: the baryonic decay $\Lambda_b \to \Lambda \tau^+ \tau^-$. 
Baryonic final states offer several experimental advantages, including reduced 
backgrounds and distinctive kinematic signatures. Furthermore, using baryons instead of mesons allows for checking the consistency of $b \to s \tau^+ \tau^-$ processes, and possibly get insights on which heavy mediator could be responsible for the aforementioned tensions.
As we will show, the LFU ratio between the tauonic and 
muonic modes, $R_\Lambda^{\tau/\mu}$,  provides a theoretically clean observable with small hadronic uncertainties. 
We present the SM prediction for this obseravable  based on lattice QCD results, explore possible NP modifications, and discuss correlations with the current anomalies in 
$b \to c \tau \nu$ and $b \to s \mu^+ \mu^-$ transitions.
We also discuss how our conclusions can be easily be adopted to the related decay mode $\Lambda_b \to p K \tau^+ \tau^-$.

This paper is organized as follows. In Section~\ref{sec:SM} we analyse the SM prediction for the $\Lambda_b \to \Lambda \tau^+ \tau^-$ and the  LFU ratio $R_\Lambda^{\tau/\mu}$
taking into account both perturbative contributions and long-distance 
effects related to the narrow charmonium resonances. As a cross-check of our description of the SM amplitude, we also briefly compare our prediction for the 
dilepton spectrum in the muon case ($\Lambda_b \to \Lambda \mu^+ \mu^-$) with current data. In Section~\ref{sec:BSM} we present a general expression for $R_\Lambda^{\tau/\mu}$ in presence of NP and discuss the enhancement expected in view of current anomalies. 
The results are summarised in the Conclusions.

\section{Standard Model predictions}
\label{sec:SM}
The effective Hamiltonian describing $b\to s \ell^+\ell^-$ transitions below the electroweak scale, both within the SM and in the general class of NP models we are interested in, reads
\begin{equation}
\label{eq:heff}
\mathcal{H}_{\mathrm{eff}}=-\frac{4 G_F}{\sqrt{2}}V_{tb}V_{ts}^* \frac{\alpha_\mathrm{em}}{4\pi} \left[ \,\sum_{i=1}^8 C_i(\mu)\mathcal{O}_i (\mu) +\sum_{i=9,10}
  C_i^\ell (\mu) \mathcal{O}_i^\ell (\mu)  \right] + \text{h.c.},
\end{equation}
where
\begin{align}
\label{eq:ops}
\mathcal{O}_1   &= 
(\bar s_L^\alpha \gamma_\mu c_L^\beta) (\bar c_L^\beta \gamma^\mu b_L^\alpha)\,,
& \mathcal{O}_2 &= 
(\bar s_L \gamma_\mu c_L)(\bar c_L \gamma^\mu b_L)\,, \nonumber \\ 
\mathcal{O}^\ell_{9} &=
(\bar{s}_L\gamma_\mu  b_L)(\bar{\ell}\gamma^\mu\ell)\,, 
&\mathcal{O}^\ell_{10}&= 
(\bar{s}_L\gamma_\mu b_L)(\bar{\ell}\gamma^\mu\gamma^5\ell)\,, \nonumber \\
 \mathcal{O}_{7} &= \frac{m_b}{e}(\bar{s}_L \sigma_{\mu\nu} b_R)F^{\mu\nu}\,. &
\end{align}
The remaining operators, $\mathcal{O}_{3-8}$, with suppressed Wilson coefficients
and vanishing tree-level matrix elements, can be found e.g.~in Ref.~\cite{Altmannshofer:2008dz}. Within the SM,
lepton flavour universality implies 
\begin{equation}
      C_{9(10)}^e \big|_{\rm SM} = 
      C_{9(10)}^\mu \big|_{\rm SM} = 
      C_{9(10)}^\tau \big|_{\rm SM} .
\end{equation}
We thus omit to indicate the lepton superscript in the Wilson coefficients
$C^\ell_9$ and $C^\ell_{10}$ in the rest of this section. We will put it back in 
Section~\ref{sec:BSM} when discussing non-universal NP effects.

Predicting observables for $\Lambda_b\to\Lambda\ell^+\ell^-$ decays involves evaluating hadronic matrix elements of the type $\langle \Lambda (p_{\Lambda})|\mathcal{O}_i|\Lambda_b(p_{\Lambda_b})\rangle$. We can distinguish three cases:
\begin{itemize}
    \item The contribution from local semileptonic $\mathcal{O}_{7,9,10}$ operators. These effects are encoded in hadronic local form-factors, scalar functions that depend on the momentum transfer $q^2 = (p_{\Lambda_b}-p_{\Lambda})^2$. To describe the $\Lambda_b\to\Lambda\ell^+\ell^-$ decays, we need ten independent form factors. For the purposes of this analysis, we use the definitions in App.~\ref{app:BR} and the numerical results from Lattice QCD computations in \cite{Detmold:2016pkz}. 
    \item The contribution from the four-quark operators $\mathcal{O}_{1-6}$. Due to Lorentz and gauge invariance, the contributions from these operators to the decay amplitude can be easily introduced as a $q^2$ dependent shift to $C_9$, defined as
    \begin{equation}
    C_9 \to C_9^{\mathrm{eff}} = C_9 + Y(q^2)~.
    \label{eq:C9eff}
    \end{equation}
    For simplicity, we can further decompose $Y(q^2)$ depending on the quark flavour:
    \begin{equation}
     Y(q^2)  =  Y_{\rm q\bar q} (q^2) + Y_{\rm c\bar{c}} (q^2) +Y_{\rm b\bar b} (q^2)\,,
    \label{eq:Ypert}
    \end{equation}
    where the first term accounts for light quarks contribution, and the last two for charm and bottom quark contributions, respectively.
    At leading order in $\alpha_s$, the function $Y(q^2)$ can be calculated perturbatively. It reads
    \begin{align}
     Y^{(0)}_{\rm q\bar q} (q^2) &= \frac{4}{3}C_3 + \frac{64}{9} C_5 + \frac{64}{27} C_6 - \frac{1}{2} h(q^2, 0) \left( C_3 +\frac{4}{3} C_4 + 16 C_5 + \frac{64}{3} C_6 \right)\,, \nonumber \\
   Y^{(0)}_{\rm c\bar c} (q^2)  &= h(q^2, m_c) \left( \frac{4}{3} C_1 + C_2 + 6 C_3 + 60 C_5 \right)\,, \nonumber \\
  Y^{(0)}_{\rm b\bar b} (q^2)    &= - \frac{1}{2} h(q^2, m_b) \left( 7 C_3 + \frac{4}{3} C_4 + 76 C_5 + \frac{64}{3} C_6 \right)\,,
\end{align}
with
\begin{equation}
    h(q^2, m) = -\frac{4}{9} \left( \ln{\frac{m^2}{\mu^2}} - \frac{2}{3} - x \right) - \frac{4}{9} (2+x) \begin{cases}
        \sqrt{x-1} \arctan{\frac{1}{\sqrt{x-1}}} \,,  & x=\frac{4 m^2}{q^2} >1\,,
        \\
        \sqrt{1-x} \left(\ln{\frac{1+\sqrt{1-x}}{\sqrt{x}}} - \frac{i \pi}{2}\right), & x=\frac{4 m^2}{q^2} \leq 1\,.
    \end{cases}
\end{equation}
Non-factorisable corrections arising at higher orders in QCD have been calculated in the literature \cite{Beneke:2001at,Asatrian:2019kbk}. Out of them, we retain only the correction to $C_7$ as in \cite{Bordone:2024hui}, while we neglect further corrections to $C_9$ given that they are numerically less important and beyond the sought for precision of this analysis. 
\item Long-distance effects from $c\bar{c}$ resonances. These are the contributions from the charmonia resonances, that are not described in the previous category. We treat them using dispersive analysis following the approach in \cite{Bordone:2024hui}. Further details are in the next section.
\end{itemize}
With this, we have all necessary information to describe the $\Lambda_b\to\Lambda\ell^+\ell^-$ decay width. We implement the differential decay width, in terms of Wilson Coefficients and hadronic local form factors, as in \app{app:BR}. The results therein have been partially crosschecked against \cite{Boer:2014kda,Blake:2017une,Bordone:2021usz}.

\subsection{Long-distance \texorpdfstring{$c\bar{c}$}{} contributions}
\begin{table}[t]
\begin{minipage}{0.35 \textwidth}
    \centering
    \begin{tabular}{|c|c|}
        \hline
        $C_1$ & $-0.291 \pm 0.009$  \\
        $C_2$ & $1.010 \pm 0.001$  \\
        $C^{\text{eff}}_7$ & -0.450 $\pm$ 0.050 \\
        $C_9$ & 4.273 $\pm$ 0.251 \\
        $C_{10}$ & -4.166 $\pm$ 0.033 \\ \hline
    \end{tabular}
    \end{minipage}
\begin{minipage}{0.65 \textwidth}
\centering
    \begin{tabular}{|c|c||c|c|}\hline
        $m_{\Lambda_b}$ & $5.620$ GeV & $m_\tau$ &  $1.776$ GeV \\ \hline
        $m_\Lambda$ & $1.116$ GeV & $m_\mu $ &   $0.106$ GeV  \\ \hline
        $m_b$ & $4.87$ GeV &  $m_e$ & $0.511\times 10^{-3}$ GeV   \\ \hline
        $|V_{tb}V^*_{ts}|$ & 0.04185 & $\tau_{\Lambda_b}$ & $ 2.2303 \times 10^{12} \text{ GeV}^{-1}$  \\ \hline
        $\alpha_{\text{em}}$ & $1/133$ & $G_F$ & $1.166 \times 10^{-5} \text{ GeV}^{-2} $  \\ \hline
    \end{tabular}
\end{minipage}
    \caption{Inputs used in the numerical analysis.  
    The Wilson Coefficients, $C_i(\mu)$, evaluated at $\mu=m_b$, are from Ref.~\cite{Bordone:2024hui}. 
    \label{tab:numinput} }
\end{table}
To include the effects from long-distance charm rescattering contributions, we closely follow the approach of  \cite{Bordone:2024hui}, where narrow charmonium resonances are described in terms of a subtracted dispersion relation. However, with respect to \cite{Bordone:2024hui}, we do not distinguish among the possible $\Lambda$ polarisation due to a lack of experimental information on the relative size of the contribution of each polarisation to the decay rate.
In the dispersive approach, we have that the $c\bar{c}$ contribution to $C_9^\mathrm{eff}$ can be rewritten as 
 \begin{align}
Y_{\rm{c\bar c}}(q^{2}) =   - \frac{4}{9} 
    \left[\frac{4}{3} C_1(\mu)  + C_2(\mu) \right] \left[ 1+
    \ln\left( \frac{m_c^2}{\mu^2} \right) \right]
 +  \frac{16 \pi^{2}}{\mathcal{F} (q^2)}  
   \sum_{V }
            \eta_V e^{i\delta_V} \frac{q^2}{m^2_{V} } A_V^\mathrm{res}(q^2)\,,
\label{eq:def_res}
\end{align} 
where the first term is the subtraction term at $q^2=0$, and each resonance $V$ is parametrised as Breit-Wigner
\begin{equation}
    A_V^{\text{res}}(q^2) = \frac{m_V \Gamma_V}{m_V^2 -q^2 - i m_V \Gamma_V} \,,
\end{equation}
with $V = J/\psi\,, \psi(2S)$, and we neglected the subleading terms proportional to parametrically small Wilson coefficients. In this context, $\mathcal{F}$ is an appropriate combination of local form factors  that, for the $\Lambda_b\to\Lambda\ell^+\ell^-$ case, reads 
\begin{align}
    \mathcal{F}^2(q^2) =  \frac{ m_{\Lambda_b}^2 }{\lambda(m_{\Lambda_b}^2, m_\Lambda^2, q^2)} \kappa_{99}(q^2) \,,
\end{align}
where
\begin{align}
    \kappa_{99} (q^2) &=  \frac{4}{3} \left[2 (|f_{\perp}|^2s_- + |g_{\perp}|^2 s_+) + \frac{(m_{\Lambda_b} + m_\Lambda)^2}{q^2} |f_+|^2 s_- + \frac{(m_{\Lambda_b} -
    m_\Lambda)^2}{q^2} |g_+|^2 s_+  \right] \,,
\end{align}
with $f_{+,\perp}$ and $g_{+,\perp}$ defined in App.~\ref{app:BR} and $s_{\pm} = (m_{\Lambda_b}\pm m_\Lambda)^2-q^2$.
The function $\kappa_{99}(q^2)$ is nothing but the combination of hadronic form factors 
appearing in the $|C_9|^2$ term in differential rate (see App.~\ref{app:BR}). 
With the $\mathcal{F}$ thus defined, we can extract $\eta_V$ for each resonance from data.
Given the definition in (\ref{eq:def_res}), the branching ratio for the resonance-mediated process $\Lambda_b\to\Lambda V\to \Lambda\ell^+\ell^-$ is
\begin{align}
    & \mathcal{B}(\Lambda_b \to \Lambda V\to \Lambda\ell^+ \ell^-) = (16\pi^2)^2 \mathcal{B}^{(0)} |\eta_V|^2\int_{4m_\ell^2}^{(m_{\Lambda_b}-m_\Lambda)^2} dq^2   |A^{\rm res}_V(q^2)|^2 \times \nonumber \\
    &\qquad\qquad\qquad\qquad\qquad\quad \times   \kappa_{99}(q^2)  \sqrt{\lambda_H(q^2) \lambda_L(q^2)} \left( 1+ \frac{2 m_l^2}{q^2}\right)  \frac{1}{|\mathcal{F}(q^2)|^2}   \bigg( \frac{q^2}{m_V^2} \bigg)^2  \nonumber \\
    & \qquad=  \frac{  (16\pi^2)^2 |\eta_V|^2 \mathcal{B}^{(0)} }{ m^2_{\Lambda_b} }  \int_{4m_\ell^2}^{(m_{\Lambda_b}-m_\Lambda)^2} dq^2  \lambda^{3/2}_H(q^2)  \lambda^{1/2}_L(q^2) 
    \frac{q^2 ( q^2 + 2 m_l^2)  }{m_V^4}  |A^{\rm res}(q^2)|^2\,, \label{eq:etaindirect}
\end{align}
where we have defined
\begin{equation}
\lambda_H(q^2)=\lambda(m_{\Lambda_b}^2\,m_\Lambda^2, q^2)\,, 
\quad 
\lambda_L(q^2) = \lambda(q^2, m_\ell^2, m_\ell^2)\,, 
\quad 
\mathcal{B}^{(0)} = \frac{G_F^2 \alpha_\text{em}^2 |V_{tb} V^*_{ts}|^2 \tau_{\Lambda_b}}{2048 \pi^5 m_{\Lambda_b}^3}\,,
\label{eq:kallenF}
\end{equation}
with $\lambda(a,b,d)=a^2+b^2+c^2-2ab-2bc-2ac$. In the narrow-width approximation, which is well motivated   for $V=J/\psi\,,\psi(2S)$, we   find
\begin{equation}
    |\eta_V|^2 = \frac{ m^2_{\Lambda_b} }{2^8 \pi^5   m_V \Gamma_V   } \times
    \frac{ \mathcal{B}(\Lambda_b \to \Lambda V )\mathcal{B}(V \to \Lambda \ell^+ \ell^- )}{  \mathcal{B}^{(0)}\lambda^{3/2}_H(m_V^2) \lambda^{1/2}_L(m_V^2) 
    (1+2 m_\ell^2/m_V^2) }\,, 
    \label{eq:eta_numerics}
\end{equation}
where we use the factorisation  $\mathcal{B}(\Lambda_b \to \Lambda V \to \Lambda \ell^+ \ell^- ) = \mathcal{B}(\Lambda_b \to \Lambda V )\mathcal{B}(V \to \Lambda \ell^+ \ell^- )$, as expected in the narrow-width approximation. Note that the $m_\ell$ dependence in the denominator of (\ref{eq:eta_numerics}) cancels the one 
in $\mathcal{B}(  V \to \ell^+ \ell^- )$.
 As a result,
in the narrow-width approximation $\eta_V$ is independent of the lepton mass.

The relevant parameters needed to obtain the values of the $\eta_V$ values from (\ref{eq:eta_numerics}) are in Table~\ref{tab:cc_resonances}. For the production of the $J/\psi$ resonance, we use the latest measurement of the LHCb collaboration \cite{lhcbcollaboration2025measurementbranchingfractionlambdab0to}, and we reconstruct the production of the $\psi(2S)$ resonance from this measurement and the world average of the ratio
\begin{equation}
    \frac{\Gamma(\Lambda_b \rightarrow \psi(2S) \Lambda)}{\Gamma(\Lambda_b\rightarrow J/\psi(1S)\Lambda)} = 0.508 \pm 0.023 \; ,
\end{equation}
in ~\cite{ParticleDataGroup:2024cfk}.
Using the values for the parameters in Tables~\ref{tab:numinput}-\ref{tab:cc_resonances}, we obtain:  $\eta_{J/\psi(1S)} = 38.0 \pm 1.8 $ and  $\eta_{\psi(2S)} = 6.65\pm 0.36$.
Concerning the phases, there are no measurements that can be employed to constrain them. Therefore, if not stated otherwise, in the following numerical analysis, we always consider them to have a flat distribution between 0 and $2\pi$.

\begin{table}[]
    \centering
    \begin{tabular}{|c|c|c|c|c|}
        \hline
        $V$ & $m_V$ & $\Gamma_V$& $\mathcal{B}(\Lambda_b\rightarrow V \Lambda)$ & $\mathcal{B}(V \rightarrow e^+ e^-)$\\ \hline
        $J/\psi(1S)$ & $3.097$ GeV & $\hphantom{0}92.6\times 10^{-6}$ GeV & $(3.34\pm 0.31) \times 10^{-4}$ & $ (5.971\pm 0.032)\times 10^{-2} $ \\ \hline
        $\psi(2S)$ & $3.686$ GeV & $293.0 \times 10^{-6}$ GeV & $(1.70 \pm 0.18 ) \times 10^{-4} $ & $(7.94 \pm 0.22) \times 10^{-3}$  \\ \hline
    \end{tabular}
    \caption{Masses and decay widths of the narrow charmonium states $J/\psi$ and $\psi(2S)$. All values are taken from \cite{ParticleDataGroup:2024cfk}, except $\mathcal{B}(\Lambda_b\to V \Lambda )$, for which we use only the latest LHCb measurement \cite{lhcbcollaboration2025measurementbranchingfractionlambdab0to}. See text for more details.
    \label{tab:cc_resonances} }
\end{table}

\subsection{Predictions for  branching ratios and 
LFU ratios}
\begin{figure}
    \centering    \includegraphics[width=0.6\linewidth]{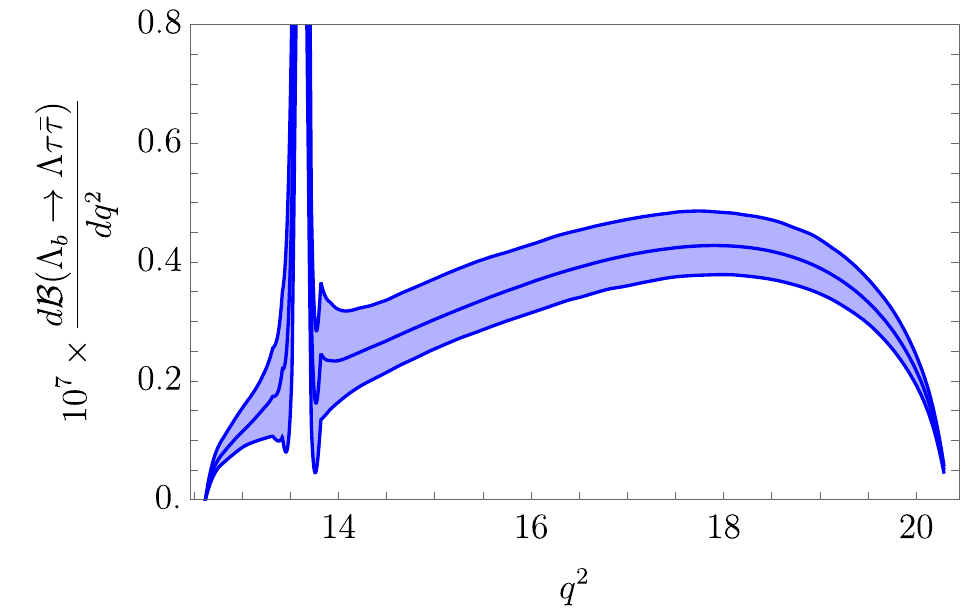}
    \caption{Predictions of the differential branching ratio of $\Lambda_b \rightarrow \Lambda \tau^+ \tau^-$ as a function of the di-lepton invariant mass $q^2$ in $\text{GeV}^2$. The band is the $1\sigma$ uncertainty region due to uncertainties from local form factors and resonance parameters. The latter includes varying the phases with a flat distribution within the interval $[0,2\pi)$.}
    \label{fig:BR}
\end{figure}

\begin{table}[t]
    \centering
    \begin{tabular}{c|c}
        sources of uncertainty & $10^7\times\mathcal{B}_{q^2>15\,\mathrm{GeV}^2}(\Lambda_b\to\Lambda\tau^+\tau^-)$  \\
        \hline
        FFs & $ 1.832^{+0.285}_{-0.218} $  \\
        $\eta_{J/\psi}, \eta_{\psi(2S)}$ &  $ 1.832^{+0.003}_{-0.003} $ \\
        FFs, $\eta_{J/\psi}, \eta_{\psi(2S)}$ & $ 1.832^{+0.285}_{-0.219} $ \\
        FFs, $\eta_{J/\psi}, \eta_{\psi(2S)}, \delta_{J/\psi}, \delta_{\psi(2S)}$ & $ 1.933^{+0.295}_{-0.228} $ \\
        $\delta_{J/\psi}, \delta_{\psi(2S)}$ & $1.933^{+0.059}_{-0.057}$
    \end{tabular}
    \caption{Uncertainties on the integrated branching ratios depending on the sources of uncertainties being considered.
    }
    \label{tab:uncertainties}
\end{table}

With the previous parametrisation of the resonances, we can now compute observables for the $\Lambda_b \to\Lambda\tau^+\tau^-$ decay. We start by studying the di-lepton branching fraction. 
Our prediction is depicted in Fig.~\ref{fig:BR}, where the blue band encodes the $1\sigma$ region accounting for the uncertainties from the local form factors and the resonance parameters. As can be seen, above $q^2\gtrsim 14\,\mathrm{GeV}^2$ the large effect coming from the $\psi(2S)$ resonance, which cannot be precisely controlled, is suppressed.  

In Table~\ref{tab:uncertainties} we present results  
 for the integrated $\mathcal{B}(\Lambda_b\to\Lambda\tau^+\tau^-)$ in the region $q^2 >15\,\mathrm{GeV}^2$. In order to provide a reliable estimate, in Table~\ref{tab:uncertainties}  we study the impact associated with various sources of uncertainty. In the first three rows we set  $\delta_{J/\psi}=\delta_{\psi(2S)}=0 $, and investigate the size of the relative uncertainties coming from the local form-factors and the parameters $\eta_{J/\psi}$ and  $\eta_{\psi(2S)}$. The uncertainty stemming from the local form factors only is in the first line, and it roughly amounts to about $14\%$. In the second line, we fix the local form factor parameters to their central values and vary only $\eta_{J/\psi}$ and  $\eta_{\psi(2S)}$. Their impact on the uncertainty is found to be subdominant, as confirmed in the third line when we vary them together with the local form factor parameters. In the fourth line, we also vary the phases $\delta_{J/\psi}$, and $\delta_{\psi(2S)}$, assuming a flat distribution for them. We notice that the central value of the branching ratio shifts, due to the fact that various choices for the phases can make the two terms in Eq.~(\ref{eq:def_res}) interfere constructively or destructively. This shift accounts for all possible combinations of phases. Finally, for completeness, in the last line we study the effect of the phase variation while setting all other parameters to their central values. We see that, in the chosen $q^2$ region, there is a residual $\sim 3\%$ error stemming from the phase variation. Taking this into account, and in the spirit of being conservative, we assume as nominal prediction: 
\begin{equation}
    \mathcal{B} (\Lambda_b\to\Lambda\tau^+\tau^-) = 1.93^{+0.30}_{-0.23}\times 10^{-7} \quad \mathrm{for} \quad q^2\geq 15\,\mathrm{GeV}^2\,.
\end{equation}
Previous determinations, which are however based on Light Cone Sum Rules or the Heavy Quark Effective Theory, are available in \cite{Aliev_2010}.

We now discuss the LFU ratio $R_\Lambda^{\tau/\mu}$, defined as 
\begin{equation}
\label{Eq:RLFU}
R_\Lambda^{\tau/\mu} = \frac{\int_{q^2_\mathrm{min}}^{(m_{\Lambda_b}-m_\Lambda)^2}\frac{d\mathcal{B}}{dq^2}(\Lambda_b\to\Lambda\tau^+\tau^-)}{\int_{q^2_\mathrm{min}}^{(m_{\Lambda_b}-m_\Lambda)^2}\frac{d\mathcal{B}}{dq^2}(\Lambda_b\to\Lambda\mu^+\mu^-)}\,.
\end{equation}
Following closely the above discussion, we obtain
\begin{equation}
(R_\Lambda^{\tau/\mu})^{\rm SM}_{[15\,{\rm GeV}^2]} = 0.526 \pm 0.033\,.
\label{eq:R_SM}
\end{equation}

The relative uncertainty in $R_\Lambda^{\tau/\mu}$ is about $7\%$, significantly smaller than that of the integrated branching fraction. This is not surprising, given most of the hadronic uncertainties cancel in the ratio between the $\Lambda_b\to\Lambda\tau^+\tau^-$ and $\Lambda_b\to\Lambda\mu^+\mu^-$ modes. However, differently from the LFU ratios between the muon and electron channels, where a relative precision around $1\%$ is achieved~\cite{Bordone:2016gaq}, $R_\Lambda^{\tau/\mu}$ has a larger uncertainty due to mass effects that are not negligible when comparing the $\tau^+\tau^-$ final state with the light leptons case. 

To a larger extent, the value of $R_\Lambda^{\tau/\mu}$ is determined 
by the available phase space, at fixed  $q^2_{\rm min}$, 
and is largely insensitive to the form factor shape. 
As we shall discuss in the next section, the SM prediction for $R_\Lambda^{\tau/\mu}$  is 
also mildly sensitive to the value of $q^2_{\rm min}$, provided 
$q^2_{\rm min} \in [15,17]~{\rm GeV}^2$, as it can be seen from the right panel of Table~\ref{tab:Rratio}. 
For this reason, the result in Eq.~\eqref{eq:R_SM}
is expected to be a reasonable estimate of analog LFU ratios in a similar kinematical range. A particularly interesting mode from the experimental point of view is the 
$\Lambda_b \to p K \ell^+\ell^-$ decay, which is 
dominated by the $\Lambda(1520)$ intermediate state.
Using the results for the $\Lambda_b\to\Lambda(1520)$ local form factors in \cite{Amhis:2022vcd} and the \texttt{EOS} software \cite{EOS:v1.0.17,EOSAuthors:2021xpv}, we estimate 
\begin{equation}
 (R_{pK}^{\tau/\mu})^{\rm SM}_{[15\,{\rm GeV}^2]}  = 0.44\pm0.03\,,
\label{eq:RpK_SM}
\end{equation}
where the central value has been obtained using only the $\Lambda(1520)$ 
intermediate state, and the uncertainty is a na\"ive estimate taking into account the contribution associated with additional intermediate states. 

\begin{figure}[t]
    \centering
\includegraphics[width=0.75\linewidth]{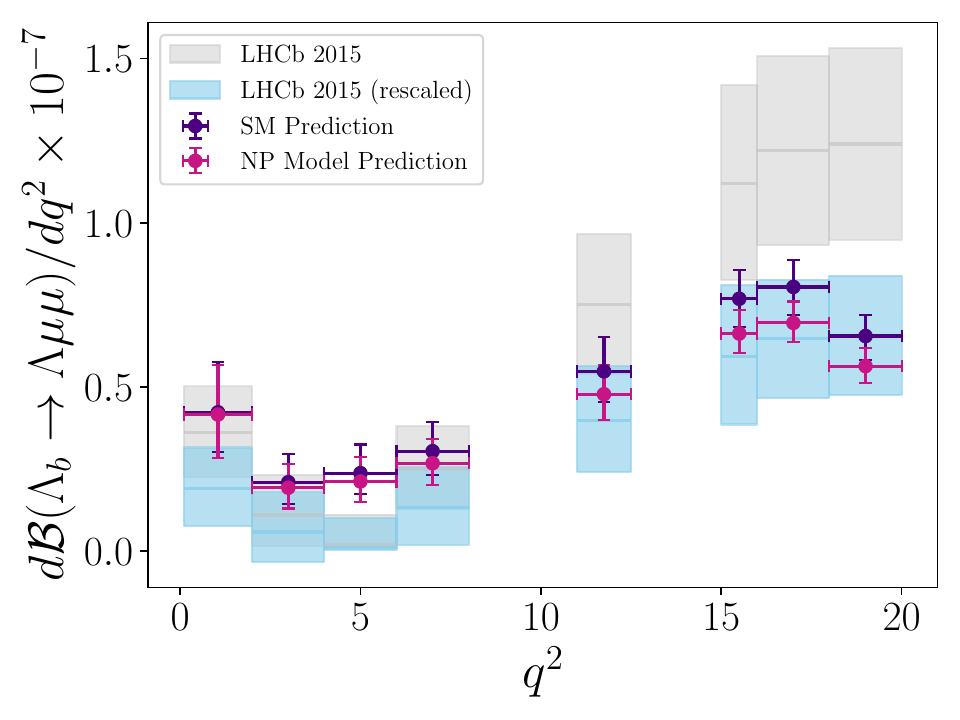}
    \caption{
    Dilepton spectrum of the $\Lambda_b\to\Lambda\mu^+\mu^-$ decay measured by LHCb.
    The gray areas are the original results reported in~\cite{LHCb:2015tgy}, while the blue ones are those obtained rescaling the normalization using the recent result for $\mathcal{B}(\Lambda_b\to \Lambda J/\psi)$ in \cite{lhcbcollaboration2025measurementbranchingfractionlambdab0to}. The crosses denote our theory prediction in the SM case  (violet) and with the NP shift to $C_9$ discussed in Section~\ref{sect:NPEFT} (purple).}
    \label{fig:mumu_predictions}
\end{figure}

Finally, we further validate our approach by comparing binned predictions for  $\mathcal{B}(\Lambda_b\to\Lambda\mu^+\mu^-)$ with the latest LHCb measurement \cite{LHCb:2015tgy}. In this work, what is actually measured is the ratio:
\begin{equation}
    \frac{\mathcal{B}(\Lambda_b\to\Lambda\mu^+\mu^-)|_{q^2_\mathrm{min}}^{q^2_\mathrm{max}}}{\mathcal{B}(\Lambda_b\to \Lambda J/\psi)}
\end{equation}
where $q^2_\mathrm{min(max)}$ is the lower(upper) extrema that define the bin length. This implies that obtaining $(\Lambda_b\to\Lambda\mu^+\mu^-)|_{q^2_\mathrm{min}}^{q^2_\mathrm{max}}$ requires an external measurement of $\mathcal{B}(\Lambda_b\to \Lambda J/\psi)$. In \cite{LHCb:2015tgy}, a then up-to-date world average combination was used. However, since then, new measurements of $\mathcal{B}(\Lambda_b\to \Lambda J/\psi)$ appeared, and, as already discussed in~\cite{Blake:2019guk}, they largely shift the reported values in \cite{LHCb:2015tgy}. We choose to rescale the results in \cite{LHCb:2015tgy} with the latest measurement from LHCb \cite{lhcbcollaboration2025measurementbranchingfractionlambdab0to}, that, at the current status, is the most precise one compared to the world average  \cite{ParticleDataGroup:2024cfk}. 
The result thus obtained is shown in Fig.~\ref{fig:mumu_predictions}. Here, in gray we report the values for the rate originally quoted in~\cite{LHCb:2015tgy}, while in light blue we plot the results of the rescaling just described. To this, we superpose our binned predictions, both in the SM case and in presence of the small NP contribution expected in the muon channel (see Section~\ref{sect:NPEFT}).  The good agreement between our predictions and the rescaled experimental data provides an important validation of the theoretical description of the hadronic part of the amplitude.

\section{Violations of LFU beyond the SM}
\label{sec:BSM}

\subsection{General parameterization of NP effects in \texorpdfstring{$R_\Lambda^{\tau/\mu}$}{}}
The NP models that we are interested in lead to modifications of the Wilson Coefficients $C^\ell_9$ and  $C^\ell_{10}$ that, in general, are lepton non-universal.
We can parameterize these effects as 
\begin{equation}
    C^\ell_9 =  C^{\rm SM}_9 + \DC^{\ell}_9\,, 
    \hspace{1cm} C^\ell_{10} = C^{\rm SM}_{10} + \DC^{\ell}_{10}\,,
\end{equation}
such that the SM limit is recovered for $\DC_{9}^\ell=\DC_{10}^\ell=0$.
The LFU ratio in Eq.~(\ref{Eq:RLFU}) is an excellent probe of NP scenarios 
where $\DC_{i}^\tau \not= \DC_{i}^\mu$. In principle, within the SM, a tiny breaking of LFU is induced by QED corrections~\cite{Bordone:2016gaq}. However, these effects do not exceed a few \% and are safely negligible for our purposes. 
The NP scenarios we are aiming to investigate are those where 
$|\DC_{9}^\tau| \gg |C_9^{\rm SM}|$ and/or 
$|\DC_{10}^\tau| \gg |C_{10}^{\rm SM}|$, leading to large 
enhancements of the $\tau^+\tau^-$ modes. 
In such limit we can also safely neglect $\DC_{9(10)}^{\mu}$
in the evaluation of $R_\Lambda^{\tau/\mu}$,
given the existing constraints on $C_{9,10}^\mu$ from
$b\to s \mu^+\mu^-$ transitions.

%
%
 
In the limit $\DC_{9,10}^{\mu}=0$, the prediction for the LFU ratio in 
Eq.~(\ref{Eq:RLFU}) can be expressed using $\DC^{\tau}_9$ and $\DC^{\tau}_{10}$ in the form
\begin{equation}
     R_\Lambda^{\tau/\mu}  =  R_{\text{SM}} \left[a_9\left(\frac{b_9}{2}+\DC^{\tau}_{9}\right)^2 + a_{10}\left(\frac{b_{10}}{2}+\DC^{\tau}_{10}\right)^2\right]\,,
\label{eq:LFU-BSM}
\end{equation}
where $R_{\text{SM}}$ is the SM value. By construction,  the numerical coefficients 
$a_i$ and $b_i$ satisfy the normalization condition
\begin{equation}
\frac{a_9 b_9^2}{4}+\frac{a_{10} b_{10}^2}{4} =1\,.
\end{equation}
Moreover, since the operator $\mathcal{O}_{10}$ does not interfere with the contributions from any other operator, $b_{10}=2  C_{10}$. As a result, taking into account also the 
normalization condition, the expression (\ref{eq:LFU-BSM})
contains three free parameters, whose numerical values 
for $q_{\rm min}^2 = 15$ GeV$^2$ are given in Table \ref{tab:Rratio}.

\begin{table}[t]
\begin{minipage}{0.5 \textwidth}
    \centering
    \begin{tabular}{c c c c}
        \toprule
        $R_{\text{SM}}$ & $a_9$ & $b_9$  \\
        \hline
        $0.526(33)$ & $0.0460(31)$ & $6.88(22)$ \\
         \midrule
         $1$ & $ -0.92 $ & $\phantom{+}0.13$ \\
         $-0.92 $ & $1$ & $ -0.51 $ \\ 
         $\phantom{+}0.13 $ & $ -0.51 $ & $1$ \\
        \bottomrule
    \end{tabular}
\end{minipage}
\begin{minipage}{0.5 \textwidth}
    \centering
    \begin{tabular}{c c }
    \toprule
        $q^2_\text{min}$ &$R_{\text{SM}}$  \\
        \midrule
        $ 15 \text{ GeV}^2$&$ 0.526 \pm 0.033  $  \\ \hline
         $ 16 \text{ GeV}^2$&$ 0.545 \pm 0.029  $ \\
         \hline
         $ 17 \text{ GeV}^2$&$ 0.561 \pm 0.024 $\\
         \hline
         $ 4 m_\tau^2$ & $ 0.463 \pm 0.038 $  \\
   \bottomrule
    \end{tabular}
\end{minipage}    
    \caption{Numerical coefficients to evaluate $R_\Lambda^{\tau/\mu}$ within and     beyond the SM via Eq.~(\ref{eq:LFU-BSM}).
    {\bf Left}: full set of coefficients for  $q_{\rm min}^2=15~{\rm GeV}^2$, 
    with correlation matrix. 
{\bf Right}: SM values for different  
choices of $q_{\rm min}^2$ 
(for $q^2_\text{min}=4 m_\tau^2$ a cut of $\pm 0.1 \text{ GeV}^2$ around $q^2=m^2_{\psi(2S)}$ is applied).}
    \label{tab:Rratio}
\end{table}

\subsection{EFT analysis}
\label{sect:NPEFT}

As already mentioned in the introduction, possible 
large NP effects in $b \to s \tau^+ \tau^-$ transitions are motivated by the 
persistent discrepancies between data and SM predictions observed in both $b \to s \mu^+ \mu^-$ and $b \to c \tau \nu$ transitions. 
In this section, we present a correlated analysis of all these effects adopting and  extending  the EFT framework developed in~\cite{Allwicher:2024ncl}.
The main hypothesis is that at some high scale $\Lambda \sim$~few TeV, the leading 
NP effects are encoded in third-generation semi-leptonic operators, with minimal breaking of the  $U(2)^5$ flavour symmetry responsible for heavy-light mixing in the left-handed quark sector. 

According to the analysis of Ref.~\cite{Allwicher:2024ncl},
scalar operators are strongly constrained by data. This allows us to restrict the attention to the following current-current  operators 
\begin{equation}
\label{eq:lagNP}
\mathcal{L}^{\rm NP}_{\mathrm{eff}} \supset  
C_{\ell q}^{+}Q_{\ell q}^{+} + C_{\ell q}^{-}Q_{\ell q}^{-} + C_{qe} Q_{qe} \,,
\end{equation}
where 
\begin{align}
Q_{\ell q}^{\pm}&=(\bar{q}_L^3 \gamma^{\mu} q_L^3)(\bar{\ell}_L^3\gamma_{\mu}\ell_L^3)  \pm ( \bar{q}_L^3 \gamma^{\mu}\sigma^a q_L^3)(\bar{\ell}_L^3\gamma_{\mu}\sigma^a\ell_L^3)\,, \nonumber \\
Q_{q e} &=(\bar{q}_L^3 \gamma^{\mu} q_L^3)(\bar\tau_R \gamma_{\mu} \tau_R)\,,
\label{eq:ops}
 \end{align}
and $\ell_L^3 = (\nu_{\tau}, \tau_L)^T$. The impact of the  $U(2)^5$ breaking in the quark sector is taken into account setting
\begin{equation}
    q_L^3 = q_L^b - \epsilon V_{ts}  q_L^s + O( V_{td} q_L^d)\,,
\end{equation}
where $q_L^{b,s,d}$ are down-aligned doublets.
The EFT framework is then described by four independent parameters: $C_{\ell q}^{\pm}$, $C_{qe}$, and $\epsilon$. The operator $C_{qe}$ has not been included in Ref.~\cite{Allwicher:2024ncl} but is worth to consider it in our analysis since it allows us to decouple NP effects in  $\DC^{\tau}_{9}$ and $\DC^{\tau}_{10}$.

\begin{figure}[t]
    \centering
    \includegraphics[width=0.48\linewidth]{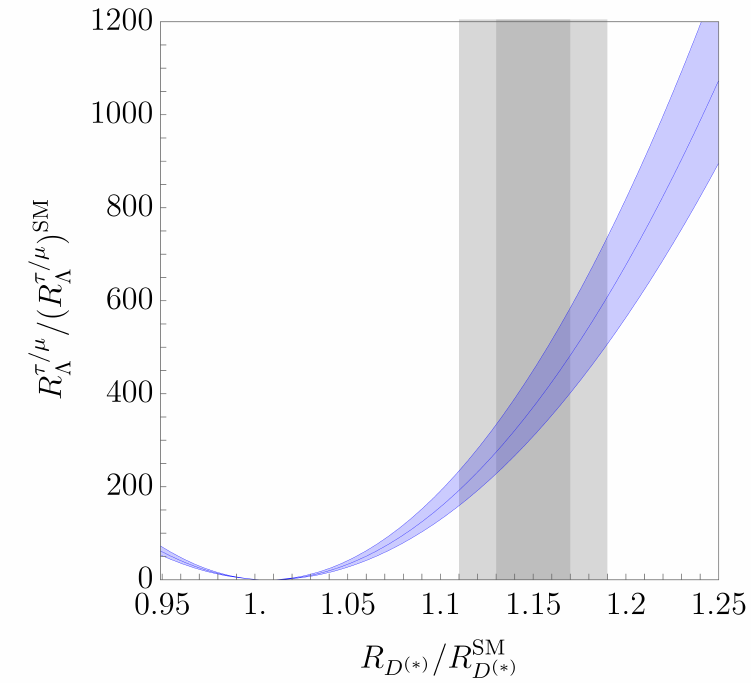}
    \hspace{2mm}
    \includegraphics[width=0.48\linewidth]{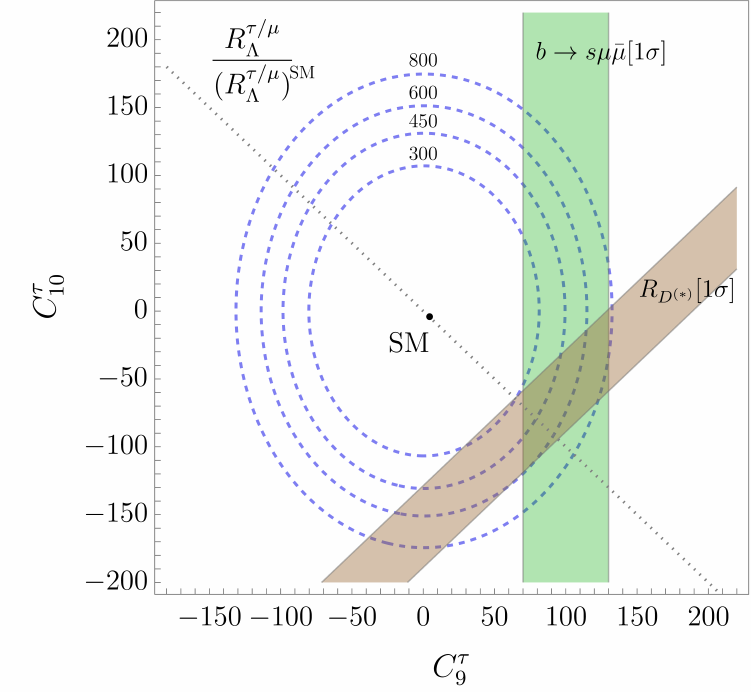}
     \caption{Predictions of $R_\Lambda^{\tau/\mu}$ in extensions of the SM with NP predominantly coupled to third-generation fermions. 
    {\bf Left:} correlation $R_\Lambda^{\tau/\mu}$ vs.~$R_{D^{(*)}}$
    in the absence of right-handed currents (blue band); both LFU ratios are 
    normalised to their SM value; 
    the dark (light) gray region indicates the experimental value of  $R_{D^{(*)}}$ at 68\% CL (98\% CL). {\bf Right:} contours in the 
    $C_9^\tau$--$C_{10}^\tau$ plane corresponding to different values of 
    $R_\Lambda^{\tau/\mu}$  normalised to the SM (dashed circles);
    the brown and green areas are those currently favored at 68\% CL 
    by $b\to s\mu\bar\mu$ and $R_{D^{(*)}}$, respectively;
    the dotted line indicates the relation
    $C_9^\tau=-C_{10}^\tau$
    expected for
    left-handed interactions.
    }
    \label{fig:BSM}
\end{figure}

Using $\mathcal{L}^{\rm NP}_{\mathrm{eff}}$ in (\ref{eq:lagNP}) the tree-level values of $\DC^{\tau}_{9}$ and $\DC^{\tau}_{10}$ reads
\begin{equation}
    \DC^{\tau}_9 = 
    -\epsilon \left(  C_{qe} + 2 C^+_{\ell q} \right)  \frac{ \pi v^2 }{ \alpha_{\rm em}}\,, \qquad
      \DC^{\tau}_{10} = -
    \epsilon \left(  C_{qe} - 2 C^+_{\ell q} \right)   \frac{  \pi v^2 }{ \alpha_{\rm em}}\,,
\end{equation}
where $v=(\sqrt{2} G_F)^{-1/2} \approx 246$~GeV and we have set $V_{tb}=1$.
The operators $Q_{\ell q}^{\pm}$ also induce a universal shift, relative to the SM, in $R_D$ and $R_{D^*}$, namely the $\tau/\ell$ ($\ell=e,\mu$) 
universality ratios of $B \to D^{(*)} \ell \nu$ transitions. The tree-level result is 
\begin{equation}
\frac{R_{D^{(*)}} }{ R^{\rm SM}_{D^{(*)}} }
\approx 1- v^2 (1+\epsilon) \left( C^+_{\ell q} - C^-_{\ell q} \right) =1- v^2 r\, \epsilon  C^+_{\ell q} \,,
\label{eq:RD}
\end{equation}
where we defined
\begin{equation}
 r =\frac{ 1 + \epsilon}{\epsilon}\left(
1 -  \frac{C^-_{\ell q}}{C^+_{\ell q} }  \right)\,.
\label{eq:rRD}
\end{equation}
The global fit of all available data performed in~\cite{Allwicher:2024ncl} setting $C_{qe}=0$, including electroweak observables,
$\mathcal{B}(B\to K^{(*)} \nu\bar\nu)$, and Drell-Yan processes at the LHC,  indicates 
$\epsilon\approx 3$ and $|C^-_{\ell q}/C^+_{\ell q}| \lesssim 0.2$. What is particularly relevant to our analysis is that the sizable value of $\epsilon$ and the smallness of  $|C^-_{\ell q}/C^+_{\ell q}|$ lead to a precise determination of the parameter $r$ in Eq.~(\ref{eq:rRD}):
$r=1.60 \pm 0.15$. As a result,  in the absence of right-handed currents,   
the NP contributions to 
 $\DC^{\tau}_{9}$, $\DC^{\tau}_{10}$, and  $R_{D^{(*)}}$ are all 
 proportional to $\epsilon  C^+_{\ell q}$ and highly correlated.  This correlation is exploited in Fig.~\ref{fig:BSM} (left), where we plot the expectation of $R_\Lambda^{\tau/\mu}$ 
 vs.~$R_{D^{(*)}}$, both normalised to their corresponding SM values, setting 
  $C_{qe}=0$. As can be seen, the current anomaly in $R_{D^{(*)}}$ favors huge 
 values $R_\Lambda^{\tau/\mu}$, as high as 
 $5\times 10^2$ or even above.

\paragraph{Combining constraints from \texorpdfstring{$b \to s \mu^+ \mu^-$}{} and \texorpdfstring{$b \to c \tau \nu$}{}.}
An independent consistency check of this NP framework is obtained by considering also information derived from $b \to s \mu^+ \mu^-$ transitions.  As pointed out in~\cite{Crivellin:2018yvo}, a huge value of $C_9^\tau$ induces, at one-loop level, a non-negligible shift in $C_9^{\mu,e}$. This effect can be estimated unambiguously, as it depends only on the ultraviolet scale $\Lambda$ at which the NP Lagrangian in Eq.~(\ref{eq:lagNP}) is defined.  The one-loop calculation
yields 
\begin{equation}
    \DC^{\mu,e}_9 =  
    -    \DC^{\tau}_9\,  
    \frac{ \alpha_{\rm em} }{ 6\pi } {\rm log}(\Lambda^2/m_\tau^2) 
    \,.
    \label{eq:Dtau}
\end{equation}
Current data on  $B\to K^{(*)}\mu^+\mu^-$ transitions~\cite{LHCb:2024onj,LHCb:2023gpo,Smith::2025} indicate a suppression of $C^{\mu,e}_9$ with respect to its SM prediction~\cite{Alguero:2022wkd,Bordone:2024hui}. 
A precise estimate of this effect is hampered by possible non-factorizable 
long-distance contributions.  Taking into account the analysis of charm rescattering presented presented in~\cite{Isidori:2024lng,Isidori:2025dkp},
we estimate $(\DC^\mu_9)^{\rm bs\mu\mu} = - 0.6 \pm 0.2$. 
 The impact of this NP contribution in the 
$\Lambda_b\to \Lambda \mu^+\mu^-$ mode is shown in Fig.~\ref{fig:mumu_predictions}.
Using this reference value and inverting Eq.~(\ref{eq:Dtau}) 
we obtain an indication of the preferred value of $\DC^\tau_9$ 
from $b\to s\mu^+\mu^-$ data. The result thus obtained, setting $\Lambda=1$~TeV is shown by the green band in Fig.~\ref{fig:BSM} (right). 

In the same plot we also display the $C_9^\tau$--$C_{10}^\tau$ region favoured by $R_{D^{(*)}}$, obtained by inverting Eq.~(\ref{eq:RD}) and expressing $\epsilon C^+_{\ell q}$ as
\begin{equation}
\epsilon C^+_{\ell q} = \frac{ \alpha_{\rm em} }{ 4\pi v^2 }
\left( \DC_{10}^\tau - \DC_{9}^\tau \right)\,.
\end{equation}
Not surprisingly, the two regions intersect in a parameter space compatible with the relation $C_{10}^\tau = -C_{9}^\tau$, which holds in the absence of right-handed currents. As also illustrated in Fig.~\ref{fig:BSM} (right), a future measurement of $R_\Lambda^{\tau/\mu}$ would provide a third, independent constraint on the $C_9^\tau$--$C_{10}^\tau$ plane and could serve as a decisive test of this NP framework.

\section{Conclusion}
\label{sec:conclusion} 

We have presented a detailed analysis of the rare baryonic decay $\Lambda_b \to \Lambda \tau^+ \tau^-$, focusing on its potential to probe new physics  coupled preferentially to third-generation fermions. Using lattice QCD results for the  $\Lambda_b \to \Lambda$ local form factors and employing a dispersive treatment of long-distance charm contributions,  we have presented an up-to-date SM prediction for ${\mathcal B}(\Lambda_b \to \Lambda \tau^+ \tau^-)$. As expected, this rate is very suppressed, in the $10^{-7}$ range, well below the current experimental sensitivity. Most importantly, we have shown that the LFU ratio $R_{\Lambda}^{\tau/\mu}$ can be predicted very precisely in the SM,
with an uncertainty below 10\%. This observable can therefore serve as a theoretically clean benchmark to test LFU in baryonic $b \to s \ell^+ \ell^-$ transitions.

Going beyond the SM, we have investigated how much $R_{\Lambda}^{\tau/\mu}$ can be enhanced in motivated 
NP scenarios in which new dynamics couple predominantly to third-generation fermions.  We have explored  the phenomenology of this class of models via a general EFT approach, relating possible lepton non-universal contributions in $b \to s \tau^+ \tau^-$, $b \to c \tau \nu$, and  $b \to s \mu^+ \mu^-$ 
decay amplitudes. As already noted in the literature, 
effective NP operators with dominant left-handed currents addressing the $R_{D^{(*)}}$ anomaly naturally imply large enhancements of 
$b \to s \tau^+ \tau^-$ rates.
 Interestingly, this scenario also implies a small suppression of the effective coefficient $C_9^\mu$ extracted from $b \to s\mu^+\mu^-$ transitions, an effect which is well consistent with current data. As we have shown, the non-trivial correlations among $R_{\Lambda}^{\tau/\mu}$, $R_{D^{(*)}}$, and $C_9^\mu$, which could be explored with future data, would provide a clear test of this NP hypothesis.

From an experimental point of view, the $\Lambda_b \to \Lambda \tau^+ \tau^-$ decay enhanced by two orders of magnitude or more, as expected in this class of NP models,
is challenging but possibly within the  reach of the LHCb experiment in the near future.  Our analysis demonstrates that, even in the absence of a direct observation, upper bounds below $10^{-4}$ would already provide useful constraints on motivated  models. The framework we have presented can also be  extended to related baryonic channels, in particular to the $\Lambda_b \to pK\tau^+\tau^-$ decay, which is particularly promising from the experimental point of view. The key aspect is measuring, or setting constraints, on ${\tau/\mu}$
LFU ratios, which are largely insensitive to hadronic uncertainties. To this purpose, we have presented the SM value for the ${\tau/\mu}$ ratio in $\Lambda_b \to pK\ell^+\ell^-$. 

In summary, rare baryonic decays with $\tau^+\tau^-$ final states offer a powerful and largely unexplored probe of possible new dynamics involving third-generation fermions. The results presented here, which include precise SM predictions and a model-independent parametrization of possible new-physics effects, offer valuable tools to fully exploit the discovery potential of future measurements.

\subsubsection*{Acknowledgements}
We thank M\'eril Reboud for helping us obtain \texttt{EOS}~predictions.
We also thank Lesya Shchutska for useful discussions that motivated us to start this work. This research was supported by 
the Swiss National Science Foundation, projects
No.~PCEFP2-194272 and 2000-1-240011.

\appendix

\section{Differential Branching Ratio}
\label{app:BR}

We parametrise the hadronic matrix elements for $\Lambda_b(p,\sLb)\to\Lambda(k,\sL)$ decays using an helicity decomposition \cite{Feldmann:2011xf,Detmold:2016pkz,Datta:2017aue,Boer:2014kda}. For the vector and the axial vector current, we have
\begin{align}
 \nonumber \langle \Lambda(k,\sL) | \bar{s} \,\gamma^\mu\, b | \Lambda_b(p,\sLb) \rangle =\,&+
 \bar{u}_\Lambda(k,\sL) \bigg[ f_0(q^2)\: (m_{\Lambda_b}-m_\Lambda)\frac{q^\mu}{q^2} \\
 \nonumber & + f_+(q^2) \frac{m_{\Lambda_b}+m_\Lambda}{s_+}\left( p^\mu + k^{\mu} - (m_{\Lambda_b}^2-m_\Lambda^2)\frac{q^\mu}{q^2}  \right) \\
 &+ f_\perp(q^2) \left(\gamma^\mu - \frac{2m_\Lambda}{s_+} p^\mu - \frac{2 m_{\Lambda_b}}{s_+} k^{ \mu} \right) \bigg] u_{\Lambda_b}(p,\sLb), \label{eq:HMEL1}\\
 \nonumber \langle \Lambda(k,\sL) | \bar{s} \,\gamma^\mu\gamma_5\, b | \Lambda_b(p,\sLb) \rangle =\,&
 -\bar{u}_\Lambda(k,\sL) \:\gamma_5 \bigg[ g_0(q^2)\: (m_{\Lambda_b}+m_\Lambda)\frac{q^\mu}{q^2} \\
 \nonumber & + g_+(q^2)\frac{m_{\Lambda_b}-m_\Lambda}{s_-}\left( p^\mu + k^{\mu} - (m_{\Lambda_b}^2-m_\Lambda^2)\frac{q^\mu}{q^2}  \right) \nonumber\\
 & + g_\perp(q^2) \left(\gamma^\mu + \frac{2m_\Lambda}{s_-} p^\mu - \frac{2 m_{\Lambda_b}}{s_-} k^{\mu} \right) \bigg]  u_{\Lambda_b}(p,\sLb)\,, \label{eq:HMEL2}
\end{align}
while for the tensor and pseudo tensor current we use:
\begin{align}
 \langle \Lambda(p^\prime,s^\prime) | \overline{s} \,i\sigma^{\mu\nu} q_\nu \, b | \Lambda_b(p,s) \rangle &=
 - \overline{u}_\Lambda(p^\prime,s^\prime) \bigg[  h_+(q^2) \frac{q^2}{s_+} \left( p^\mu + p^{\prime \mu} - (m_{\Lambda_b}^2-m_{\Lambda}^2)\frac{q^\mu}{q^2} \right) \\
 & + h_\perp(q^2)\, (m_{\Lambda_b}+m_\Lambda) \left( \gamma^\mu -  \frac{2  m_\Lambda}{s_+} \, p^\mu - \frac{2m_{\Lambda_b}}{s_+} \, p^{\prime \mu}   \right) \bigg] u_{\Lambda_b}(p,s), \nonumber  \\
  \langle \Lambda(p^\prime,s^\prime)| \overline{s} \, i\sigma^{\mu\nu}q_\nu \gamma_5  \, b|\Lambda_b(p,s)\rangle &=
 -\overline{u}_{\Lambda}(p^\prime,s^\prime) \, \gamma_5 \bigg[   \widetilde{h}_+(q^2) \, \frac{q^2}{s_-} \left( p^\mu + p^{\prime \mu} -  (m_{\Lambda_b}^2-m_{\Lambda}^2) \frac{q^\mu}{q^2} \right) \\
 & + \widetilde{h}_\perp(q^2)\,  (m_{\Lambda_b}-m_\Lambda) \left( \gamma^\mu +  \frac{2 m_\Lambda}{s_-} \, p^\mu - \frac{2 m_{\Lambda_b}}{s_-} \, p^{\prime \mu}  \right) \bigg]  u_{\Lambda_b}(p,s), \nonumber
\end{align}
with $q=p-k$, $s_\pm =(m_{\Lambda_b} \pm m_\Lambda)^2-q^2$, and $\sLb$ and $\sL$ are the spin of the $\Lambda_b$ and $\Lambda$ baryons, respectively. 
With these definitions, we write down the differential branching fraction:
\begin{equation}
    \frac{d \mathcal{B}(\Lambda_b\to \Lambda \ell^+\ell^-)}{dq^2 d\cos\theta} = \mathcal{B}^{(0)}  \frac{ \sqrt{\lambda_H(q^2) \lambda_L(q^2)}}{q^2}
    [A(q^2) + B(q^2) \cos\theta+C(q^2) \cos^2\theta] ,
\end{equation}
with $\lambda_{H,L}(q^2)$ and $\mathcal{B}^{(0)}$ defined as in Eq.~(\ref{eq:kallenF}) and 
where  the angular coefficients are
\begin{align}
    A(q^2) &= |C_{10}|^2 \frac{4m_\ell^2}{q^2} \bigg( M_-^2 s_+ |f_0|^2 + M_+^2 s_- |g_0|^2 \bigg) \nonumber \\ &+\bigg( |C_9|^2 (4m_\ell^2 +q^2) - |C_{10}^2| q_+ \bigg) \bigg(   s_- |f_\perp|^2 + s_+ |g_\perp|^2 \bigg)   \nonumber \\ 
    &   +  \bigg( |C_9|^2 - \frac{q_+}{q^2} |C_{10}|^2 \bigg)\bigg( s_-M_+^2 |f_+|^2 + s_+ M_-^2 |g_+|^2 \bigg)  \nonumber \\ 
    &+ |C_7|^2 \frac{4m_b^2}{q^4} \bigg( s_-(q^4 |h_+|^2 +M_+^2 (q^2+4m_\ell^2) |h_\perp|^2) + s_+ (q^4 |\tilde h_+|^2 +M_-^2 (q^2 +4m_\ell^2) |\tilde h_\perp|^2 ) \bigg)  \nonumber \\
    &+ \frac{4m_b}{q^2} \bigg( M_+ s_- (q^2 \Re(C_7 C_9^* h_+ f_+^*)  +(q^2+4m_\ell^2)\Re(C_7 C_9^* h_\perp f_\perp^*)) \nonumber \\ 
    &+ M_- s_+ ( q^2 \Re(C_7 C_9^* \tilde h_+ g_+^*) + (q^2+4m_\ell^2) \Re(C_7 C_9^* \tilde h_\perp g_\perp^*)) \bigg) ,
\end{align}

\begin{align}
    B(q^2) &= - \sqrt{-\frac{q_+ \lambda_H}{q^2}} \Bigg( 8 m_b  \Big( M_- \Re( C_7 C_{10}^* \tilde h_\perp f_\perp^*)  + M_+ \Re( C_7 C_{10}^* h_\perp g_\perp^* )\Big) + 8 q^2 \Re( C_9 C_{10}^*) \Re( f_\perp g_\perp^*)  \Bigg) ,
\end{align}

\begin{align}
    C(q^2) &= (|C_9|^2 + |C_{10}|^2) \frac{q_+}{q^2} \bigg( s_-(M_+^2 |f_+|^2-q^2 |f_\perp|^2) + s_+(M_-^2 |g_+|^2 -q^2 |g_\perp^2| )\bigg) \nonumber \\
    &+|C_7|^2 \frac{4m_b^2 q_+}{q^4} \bigg( s_-(q^2 |h_+|^2 -M_+^2 |h_\perp|^2) + s_+ (q^2 |\tilde h_+|^2 -M_-^2 |\tilde h_\perp|^2 ) \bigg) \nonumber \\
    &+ 4m_b \frac{q_+}{q^2} \bigg( M_+ s_- \Re(C_7 C_9^* (h_+ f_+^*- h_\perp f_\perp^*)) + M_- s_+ \Re(C_7 C_9^*(\tilde h_+ g_+^*- \tilde h_\perp g_\perp^*))   \bigg),
\end{align}
where we define
    $q_+ = 4 m_\ell^2-q^2$ and  $M_\pm = (m_{\Lambda_b} \pm m_\Lambda)$.


\bibliographystyle{JHEP} 
\bibliography{refs.bib}
\end{document}